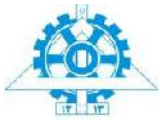
College of Engineering

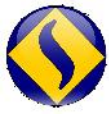
Society of Manufacturing Engineering of Iran

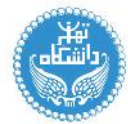
University of Tehran

# Design, Analysis, and Simulation of a Pipe-Welding Robot with Fixed Plinth

A. Emami[*], S. Khaleghian and M.J. Mahjoob

*School of Mechanical Engineering, College of Engineering, University of Tehran, Tehran, Iran*

**Abstract**

Industrial requirements concerning the increased efficiency and high rate of manufacturing result in the development of manufacturer robots, and a vast group of these types of robots is used for welding. This study presented the design, analysis, and simulation of a pipe-welding robot with fixed plinth for a constant circular welding around the pipes. Design of a welding robot capable of keeping the electrode orientation, welding speed, and distance between electrode and pipe surface constant can improve the quality of welding; thus, a five-linked articulated robot was designed for this purpose. Solving of direct and diverse kinematics and dynamics' equations of the robot was done by means of Matlab software. The robot was also simulated using a program written in Matlab and the diagrams of angles, velocities, and accelerations of all the arms, and the applied force and torque of each arm required for drive the mechanism were obtained.

## 1. Introduction

Commercial and industrial robots are now in widespread use performing jobs more cheaply with greater accuracy and reliability than humans; as a result, robotics is an essential component in any modern manufacturing environment. As factories increase their use of robots, the number of robotics related jobs grow and have been observed to be on a steady rise. Nowadays, industrial robots have many applications in manufacturing such as welding, painting, assembly, testing, etc. all accomplished with high endurance, speed, and precision.

Since welding is an important process in manufacturing mechanical equipment and needs skill and high precision, development of welding robots can benefit manufacturer by improving welding quality, accelerating the welding process, and increasing its accuracy[1]. Robot welding is a relatively new application of robotics, even though robots were first introduced into US industry during the 1960s. The use of robots in welding did not take off until the 1980s, when the automotive industry began using robots extensively for spot welding. Since then, both the number of robots used in industry and the number of their applications has grown greatly [2]. In 2005, more than 120,000 robots were in use in North American industry, about half of them for welding [3]. Robot arc welding has begun growing quickly just recently, and already it commands about 20% of industrial robot applications. Many researchers also worked on motion navigation and control of arc welding robot such as Zhenyu Liu et al [4], S Murakami et al [5], and Doyoung Chang et al [6].

While large amount of stainless steel pipes would be applied in the pipeline transport system, if the pipes and tubes are welded manually, there would be a lot of the repeated work accompanied with the extremely low efficiency, and the welding quality could not guaranteed neither. Therefore, the best choice to finish the work is using a robotic welding system [7]. In this study, a robotic system with a fixed plinth to be employed in welding the pipes for transportation of oil, gas, and refined products between cities and countries is introduced. This mechanism contains one prismatic and five revolute joints. The prismatic joint located on the ground is used for adjusting the height of the plinth of robot according to centerline of the pipe. After adjusting the first revolute joint in the same height of pipe centerline, the plinth becomes fixed and welding process begins. The robotic configuration was design so that the robotic arms can be adjusted to a wide range of pipes' diameter.

In the motion study of the robot, direct and invert kinematics and dynamics' equations were solved utilizing Matlab software, and the results of this solution were utilized for simulating the robot. Simulation codes were also written in Matlab. The outcomes of this simulation were the diagrams of angles, velocities, and accelerations of all the arms, and the applied force and torque of each arm required for drive the mechanism. These diagrams could help to select the suitable step motors for operation of the welding robot and guiding the tip of electrode on the circular path around the pipe.

---
[*]Corresponding author. Tel.: +98 21 22717188, Fax: +98 21 22717188
E-mail address: emami_anahita@yahoo.com

## 2. Robot Geometry

The most commonly used robot configurations in industry are articulated robots, SCARA robots and Cartesian coordinate robots. Since in pipe welding, the robot should have the ability to track the circular path in constant motion without intersecting the path to avoid collision of robot arms with pip surface, an articulated robot is more suitable than other robot configurations. The schematic robot was shown in Fig. 1. All the coordinate frames and axes used for analysis of robot are shown in this figure.

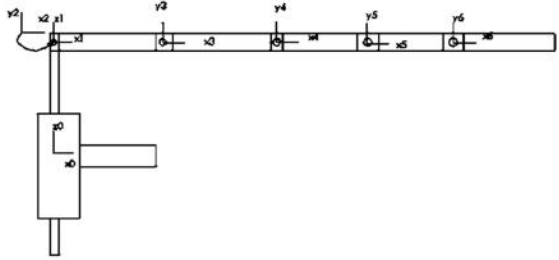

Fig. 1. Schematic robot with its coordinate frames

All the geometric parameters of this robot are shown in Table 1. In this table , a, d, and are link twist, link length, joint offset, and joint angle respectively.

Table 1 geometric parameters of robot

| $i$ | $\alpha_{i-1}$ | $a_{i-1}$ | $d_i$ | $\theta_i$ |
|---|---|---|---|---|
| 1 | 0 | 0 | L1 | 0 |
| 2 | 90 | L2 | 0 | $\theta_2$ |
| 3 | 0 | L3 | 0 | $\theta_3$ |
| 4 | 0 | L4 | 0 | $\theta_4$ |
| 5 | 0 | L5 | 0 | $\theta_5$ |
| 6 | 0 | L6 | 0 | $\theta_6$ |

## 3. Motion Study

The study of motion can be divided into kinematics and dynamics. Direct kinematics refers to the calculation of end effector position, orientation, velocity, and acceleration when the corresponding joint values are known. Inverse kinematics refers to the opposite case in which required joint values are calculated for given end effector values, as done in path planning. Some special aspects of kinematics include handling of redundancy (different possibilities of performing the same movement), collision avoidance, and singularity avoidance. Once all relevant positions, velocities, and accelerations have been calculated using kinematics, methods from the field of dynamics are used to study the effect of forces upon these movements. Direct dynamics refers to the calculation of accelerations in the robot once the applied forces are known. Direct dynamics is used in computer simulations of the robot. Inverse dynamics refers to the calculation of the actuator forces necessary to create prescribed end effector acceleration. This information can be used to improve the control algorithms of a robot. [8]

### 3.1. Direct Kinematics

Transformation matrix of coordinate frame number $i$ to $i$-$1$ defines as:

$$^{i-1}_{i}T = \begin{vmatrix} C\theta_i & -S\theta_i & 0 & a_{i-1} \\ S\theta_i C\alpha_{i-1} & C\theta_i C\alpha_{i-1} & -S\alpha_{i-1} & -S\alpha_{i-1}d_i \\ S\theta_i S\alpha_{i-1} & C\theta_i S\alpha_{i-1} & C\alpha_{i-1} & C\alpha_{i-1}d_i \\ 0 & 0 & 0 & 1 \end{vmatrix}$$
(1)

Where $C$ stands for cosine and $S$ stands for sine.

Transformation matrix of last coordinate frame to base frame can be achieved from following equation:

$$^{0}_{6}T = ^{0}_{1}T\,^{1}_{2}T\,^{2}_{3}T\,^{3}_{4}T\,^{4}_{5}T\,^{5}_{6}T$$
(2)

Using robotic parameters of each link in Eq. (1) leads to six matrices. Substitution of these matrices in Eq. (2) leads to the main kinematics equations' matrix of the robot which defines as:

$$^{0}_{6}T = \begin{vmatrix} C_{65432} & -S_{65432} & 0 & L_5C_{5432}+L_4C_{432}+L_3C_{32}+L_2C_2 \\ 0 & 0 & -1 & 0 \\ S_{65432} & C_{65432} & 0 & L_5S_{5432}+L_4S_{432}+L_3S_{32}+L_2S_2+L_1 \\ 0 & 0 & 0 & 1 \end{vmatrix}$$
(3)

Where $C_{65} = \cos(\theta_6 + \theta_5)$, $S_{65} = \sin(\theta_6 + \theta_5)$, and so forth.

### 3.2. Inverse Kinematics

The transformation matrix of tool frame to base frame is obtained from kinematics calculation which defines as:

$$^{0}_{t}T = \begin{vmatrix} C_{65432} & -S_{65432} & 0 & L_6C_{65432}+L_5C_{5432}+L_4C_{432}+L_3C_{32}+L_2C_2 \\ 0 & 0 & -1 & 0 \\ S_{65432} & C_{65432} & 0 & L_6S_{65432}+L_5S_{5432}+L_4S_{432}+L_3S_{32}+L_2S_2+L_1 \\ 0 & 0 & 0 & 1 \end{vmatrix}$$
(4)

Matrix $^{0}_{t}T$ is also equal with following matrix which defines the position of the tip of electrode on the circular path around the pipe:

$$^0_tT = \begin{matrix} r_{11} & r_{13} & 0 & p_x \\ 0 & 0 & -1 & 0 \\ r_{31} & r_{32} & 0 & p_z \\ 0 & 0 & 0 & 1 \end{matrix} \quad (5)$$

Equality of components of matrix (4) and (5) leads to following equations:

$$\theta_6 + \theta_5 + \theta_4 + \theta_3 + \theta_2 = \text{Atan2}(r_{31}, r_{11}) \quad (6)$$
$$p_x - L_6 C_{65432} - L_3 C_{32} - L_2 C_2 = L_5 C_{5432} + L_4 C_{432} \quad (7)$$
$$p_z - L_1 - L_6 S_{65432} - L_3 S_{32} - L_2 S_2 = L_5 S_{5432} + L_4 S_{432} \quad (8)$$

Since three independent equations could be extracted from diverse kinematics equations and five joint angles are their variables, $\theta_2$ and $\theta_3$ were evaluated in order to find a solution for diverse kinematics equations. Then, squaring Eq. (7) and Eq. (8) and adding them with each other leads to the following equation:

$$(p_x - L_6 C_{65432} - L_3 C_{32} - L_2 C_2)^2 + (p_z - L_1 - L_6 S_{65432} - L_3 S_{32} - L_2 S_2)^2 = K \quad (9)$$

Where $K = L_5^2 + L_4^2 + 2L_2 L_5 C_5$
Following equations are extracted from Eq. (9):

$$C_5 = \frac{K - L_5^2 - L_4^2}{2L_4 L_5} \quad (10)$$

$$S_5 = \pm \sqrt{1 - C_5^2} \quad (11)$$

Using Eq. (10) and Eq. (11) following equation was extracted for $\theta_5$:
$$\theta_5 = \text{Atan2}(S_5, C_5) \quad (12)$$

By similar calculation, following equations were extracted:

$$S_{234} = \frac{L_4 + C_5 L_5 \, B - S_5 L_5 A}{L_5^2 + L_4^2 + 2L_2 L_5 C_5} \quad (13)$$

$$C_{234} = \frac{L_5 + C_5 L_4 \, -C_5 B + S_5 A}{L_4 S_5} \quad (14)$$

Where $A = p_x - L_6 C_{65432} - L_3 C_{32} - L_2 C_2$ and
$B = p_z - L_1 - L_6 S_{65432} - L_3 S_{32} - L_2 S_2$

Using Eq. (13) and Eq. (14) following equation was extracted for $\theta_4$:
$$\theta_4 = \text{Atan2}(S_{234}, S_{234}) - \theta_3 - \theta_2 \quad (15)$$
Using Eq. (6), Eq. (12), and Eq. (15) following equation was also extracted for $\theta_6$:

$$\theta_6 = \text{Atan2}(r_{31}, r_{11}) - \text{Atan2}(\pm\sqrt{1 - C_5^2}, C_5)$$
$$- \text{Atan2}(S_{234}, S_{234}) \quad (16)$$

Solving Eq. (12), Eq. (15), and Eq. (16) in Matlab software for given end effector values gave joint values. In this study joint values for a circular path which was pipe perimeter were calculated.

### 3.3. Dynamics

Dynamic analysis of robot was using algorithm of iterative relations which defines as:

$$^{i+1}\check{S}_{i+1} = \,^{i+1}_i R \,^i\check{S}_i + \dot{\theta}_{i+1} \,^{i+1}\hat{Z}_{i+1}$$
$$^{i+1}\dot{\check{S}} = \,^{i+1}_i R \,^i\dot{\check{S}}_i + \,^{i+1}_i R \,^i\check{S}_i \times \dot{\theta}_{i+1} \,^{i+1}\hat{Z}_{i+1} + \ddot{\theta}_{i+1} \,^{i+1}\hat{Z}_{i+1}$$
$$^{i+1}\tilde{v}_{i+1} = \,^{i+1}_i R(\check{S}_i \times \,^i P_{i+1} + \,^i\check{S}_i \times (\,^i\check{S}_i \times \,^i P_{i+1}) + \,^i\tilde{v}_i)$$
$$^{i+1}\tilde{v}_{c_{i+1}} = \,^{i+1}\dot{\check{S}}_{i+1} \times \,^{i+1}P_{c_{i+1}} + \,^{i+1}\check{S}_{i+1} \times (\,^{i+1}\check{S}_{i+1} \times \,^{i+1}P_{c_{i+1}}) + \,^{i+1}\tilde{v}_{i+1}$$
$$^{i+1}F_{i+1} = m_{i+1} \,^{i+1}\tilde{v}_{c_{i+1}}$$
$$^{i+1}N_{i+1} = \,^{c_{i+1}}I_{i+1} \,^{i+1}\dot{\check{S}}_{i+1} + \,^{i+1}\check{S}_{i+1} \times \,^{c_{i+1}}I_{i+1} \,^{i+1}\check{S}_{i+1}$$

(17)

Where $i$ varies from 0 to 5.
Starting the algorithm from base frame gives the velocity, acceleration, total forces, and torques of center mass of single link in each step. After the last step of algorithm (17), all the joint actuator forces and actuator torques can be found using the results of algorithm (17) in following algorithm:

$$^i f_i = \,^i_{i+1} R \,^{i+1}f_{i+1} + \,^i F_i$$
$$^i n_i = \,^i N_i + \,^i_{i+1} R \,^{i+1}n_{i+1} + \,^i P_{c_i} \times \,^i F_i + \,^i P_{i+1} \times \,^i_{i+1} R \,^{i+1}f_{i+1}$$
$$\ddagger_i = \,^i n_i^T \,^i \hat{Z}_i$$

(18)

Where $i$ varies from 6 to 1.

### 4. Simulation

A sample of this pipe-welding robot was simulated for a pipe with 1-meter diameter. All the robotic arms were made of Plexiglas with density of 1.19grams/cubic centimeter, and their widths and thicknesses were assumed 30mm and 4mm respectively. The lengths of all the moving arms are shown in Table 2. (Note: The length of first link which is the plinth of robot should be adjusted according to the height of pipe. In the calculations of simulation, the Length of first arm was assumed 3.45m)

Table 2 Lengths of robot arms

| | |
|---|---|
| L2 | 3.22m |
| L3 | 2.99m |
| L4 | 2.76m |
| L5 | 2.53m |
| L6 | 2.3m |

### 4.1. Motion simulation coding

The simulation code for the sample robot was written in Matlab. It included a sub function which avoided any intersection of robot arms with the circular path of welding on external surface of the pipe. Since the electrode had to be located in the end of last arm and its direction should be kept perpendicular to the pipe surface (in the direction of the pipe radius), the code included another sub function which kept the end of last arm on the path with constant motion and no sudden change in direction and orientation. The simulation program presented the animated result of robot motion for a complete welding of the pipe. The image of a moment in animation is shown in Fig. 2.

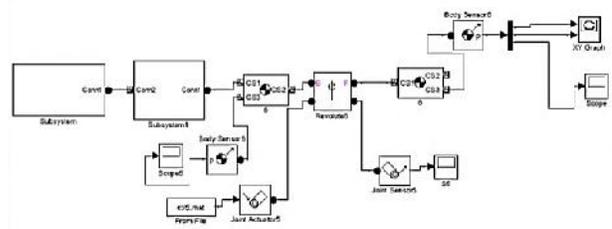

Fig. 4 Main body blocks of Sim-Mechanic

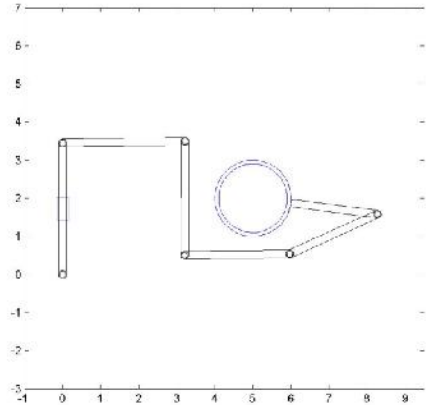

Fig. 3 Simulated pipe-welding robot

The simulation code also calculated all the angles, angular velocities, and angular accelerations of robot arms in a complete welding path around the pipe and it gave them in separate diagrams for each arm. Using these results in dynamics equations gave the required external forces and torques which should be applied by suitable step motors.

### 4.2. Invert kinematics simulation by Sim-Mechanic

The purpose of invert kinematics simulation using Sim-Mechanic of Simulink in Matlab software was the verification of invert kinematic calculations which had been done previously. All the results of invert kinematics calculations employed as inputs of Sim-Mechanic's blocks which are shown in Fig. 3, Fig. 4, and fig.5. The output of the system was the path of last arm's tip which was sketched by XY Graph. The outcome sketch was a circle with the same diameter of the pipe and it showed the accuracy of invert kinematics solutions.

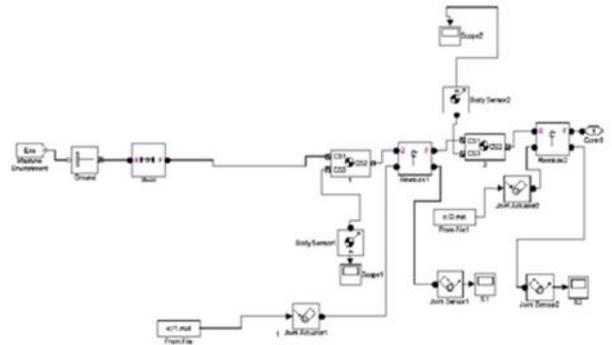

Fig. 2 Subsystem blocks

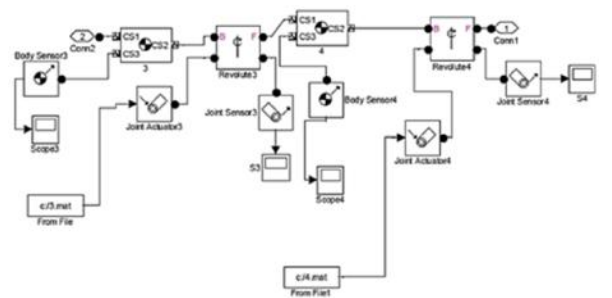

Fig. 5 Subsystem1 blocks

### 5. Results

All the results of direct kinematics, inverse kinematics, and dynamics solution came in six diagrams for each robot arm. These diagrams can be utilized to select the proper step motor for each link. They also can be used for programing the required controllers of the robot. The diagrams of last link are shown in Fig. 6 to Fig. 11.

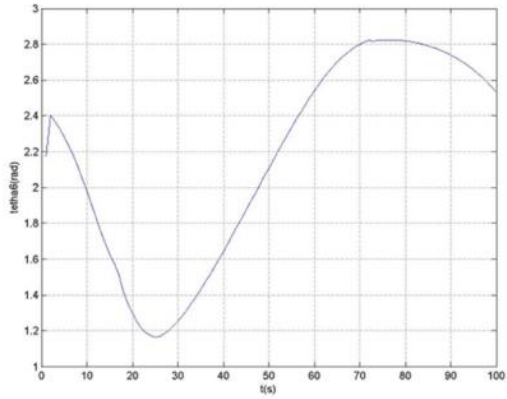

Fig. 6 angular position of last joint $(\theta_6)$

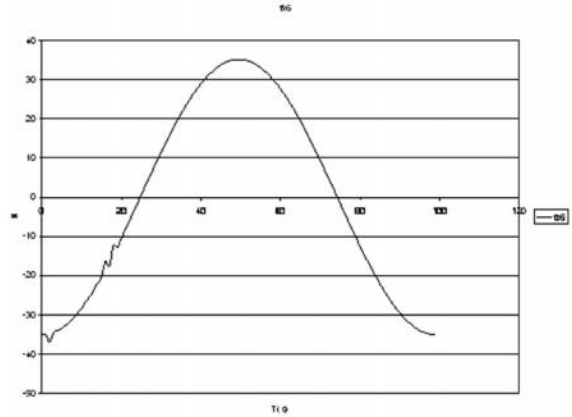

Fig. 9 Horizontal actuator force of last joint $(F_{x_6})$

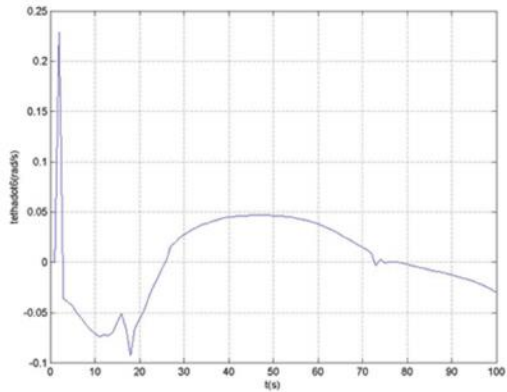

Fig. 7 angular velocity of last joint $(\dot{\theta}_6)$

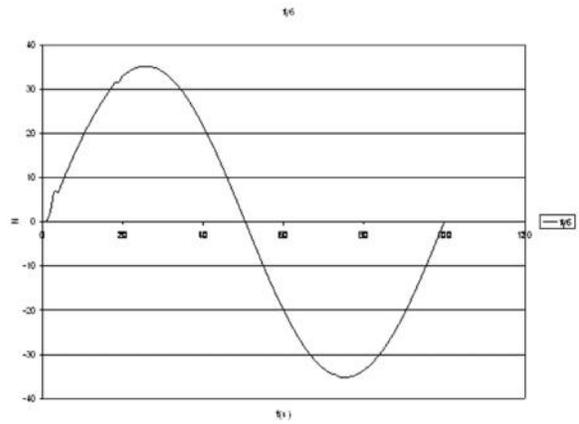

Fig. 10 Vertical actuator force of last joint $(F_{y_6})$

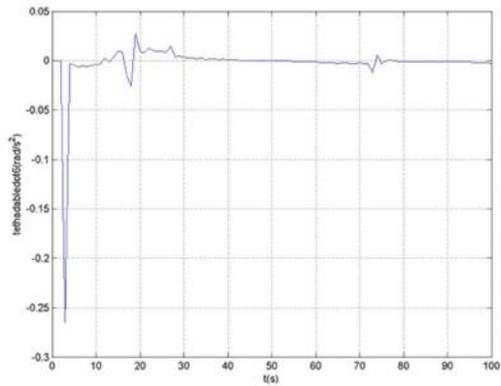

Fig. 8 angular acceleration of last joint $(\ddot{\theta}_6)$

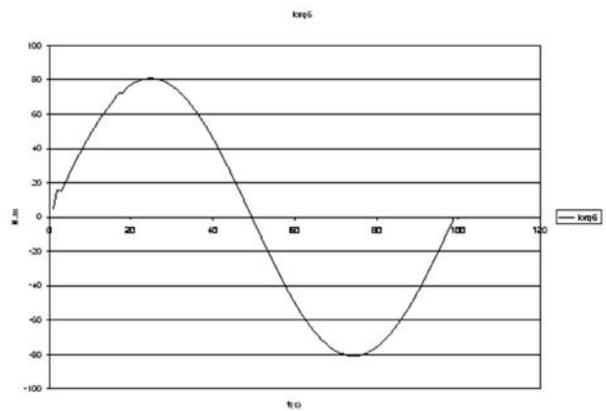

Fig. 11 Actuator torque of last joint $(T_6)$

**6. Conclusions**

In this study, a type of pipe-welding robot with fixed plinth was designed. The advantage of this robot is its ability to operate in different height and for vast range of pipes with different diameters. Moreover, the written code for the sample robot can be used for robots with different dimensions just by changing the length parameters used in the first line of the code.

**References**

[1] Mustafa Suphi Erden, Bobby Mari "Assisting manual welding with robot" *Article Robotics and Computer-Integrated Manufacturing*, Volume 27, Issue 4, August 2011, Pages 818-828

[2] R.D. Schraft "The industrial robot in a flexible manufacturing system: State of the art and prospects" *Robotics*, Volume 2, Issue 3, September 1986, Pages 237-247

[3] Cary, Howard B. and Scott C. Helzer "Modern Welding Technology", Upper Saddle River, New Jersey: Pearson Education. Page 316. ISBN 0-13-113029-3.,2005.

[4] Zhenyu Liu, Wanghui Bu, Jianrong Tan "Motion navigation for arc welding robots based on feature mapping in a simulation environment" *Robotics and Computer-Integrated Manufacturing*, Volume 26, Issue 2, April 2010, Pages 137-144

[5] S Murakami, F Takemoto, H Fujimura, E Ide "Weld-line tracking control of arc welding robot using fuzzy logic controller" *Article Fuzzy Sets and Systems*, Volume 32, Issue 2, 11 September 1989, Pages 221-237

[6] Doyoung Chang, Donghoon Son, Jungwoo Lee, Donghun Lee, Tae-wan Kim, Kyu-Yeul Lee, Jongwon Kim "A new seam-tracking algorithm through characteristic-point detection for a portable welding robot" *Robotics and Computer-Integrated Manufacturing*, Volume 28, Issue 1, February 2012, Pages 1-13

[7] Werner Neubauer "Locomotion with articulated legs in pipes or ducts" *Robotics and Autonomous Systems* Volume 11, Issues 3-4, December 1993, Pages 163-169

[8] J.J. Craig, "Introduction to robotics: Mechanism and Control", 2nd Ed, Addison-Wesley Publishing Co., Reading, MA, 1989.